\documentclass[aps,notitlepage,12pt,showkeys,tightenlines]{revtex4}

\usepackage{amsmath,amssymb,amsfonts}
\usepackage[mathcal]{euscript}
\usepackage{graphicx}
\usepackage{psfrag}

\numberwithin{equation}{section}

\newcommand{\commentout}[1]{}

\newcommand{\etal}{\textit{et~al.}}

\newcommand{\mathnotation}[2]{\newcommand{#1}{\ensuremath{#2}}}
\newcommand{\Tr}[1]{\overset{\times}{#1}}

\newcommand{\Trless}[1]{\overset{\circ}{#1}}
\newcommand{\Trlessw}[1]{\overset{\circ}{\overline{#1}}}
\newcommand{\transp}[1]{\widetilde{#1}}
\newcommand{\Dt}[1]{D_\time#1}

\newcommand{\Order}[1]{\mathcal{O}\l(#1\r)}

\renewcommand{\l}{\left}			
\renewcommand{\r}{\right}			
\mathnotation{\pd}{\partial}			
\mathnotation{\pdt}{\partial_\time}		
\mathnotation{\ee}{{\mathrm e}}			
\mathnotation{\grad}{{\boldsymbol{\nabla}}}		
\mathnotation{\ldef}{\mathrel{\raisebox{.069ex}{:}\!\!=}}
\mathnotation{\rdef}{\mathrel{=\!\!\raisebox{.069ex}{:}}}
\mathnotation{\half}{\tfrac{1}{2}}		
\mathnotation{\third}{\tfrac{1}{3}}		
\mathnotation{\threehalf}{\tfrac{3}{2}}		
\mathnotation{\curl}{\grad\times}		
\renewcommand{\div}{\grad\cdot}			
\mathnotation{\lapl}{\nabla^2}			
\mathnotation{\blapl}{\nabla^4}			
\mathnotation{\pheq}{&\phantom{=}}		
\mathnotation{\dint}{\,{d}}			
\mathnotation{\solidang}{\Omega}		

\renewcommand{\time}{t}				
\mathnotation{\x}{x}				
\mathnotation{\xv}{{\mathbf{\x}}}		
\mathnotation{\vc}{v}				
\mathnotation{\vv}{{\mathbf{\vc}}}		
\mathnotation{\uc}{u}				
\mathnotation{\uv}{{\mathbf{\uc}}}		
\mathnotation{\cc}{c}				
\mathnotation{\cv}{{\mathbf{\cc}}}		
\mathnotation{\RTv}{\alpha}			
\mathnotation{\cs}{a}				
\mathnotation{\cnu}{w}				
\mathnotation{\cnuv}{\boldsymbol{w}}	
\mathnotation{\ku}{\hat{k}}			
\mathnotation{\kuv}{{\mathbf{\ku}}}		
\mathnotation{\expar}{\epsilon}			
\mathnotation{\relt}{\tau}			
\mathnotation{\relti}{\relt_1}			
\mathnotation{\reltii}{\relt_2}			
\mathnotation{\Macrotime}{\mathfrak{T}}		
\mathnotation{\sT}{s}				
\mathnotation{\gam}{\gamma}			
\mathnotation{\Liouv}{{\mathcal{D}}}		
\mathnotation{\Coll}{{\mathcal{C}}}		
\mathnotation{\Collt}{\widetilde{\Coll}}	
\mathnotation{\accel}{\mathbf{a}}		
\mathnotation{\accelv}{{\mathbf{\accel}}}	
\mathnotation{\E}{E}				
\mathnotation{\Ev}{\mathbf{\E}}			
\mathnotation{\B}{H}				
\mathnotation{\Bv}{\mathbf{\B}}			
\mathnotation{\elc}{e}				
\mathnotation{\mass}{m}				
\mathnotation{\dens}{\rho}			
\mathnotation{\ndens}{n}			
\mathnotation{\Temp}{T}				
\mathnotation{\ddens}{\phi}
\mathnotation{\dTemp}{\theta}
\mathnotation{\dpres}{\varpi}
\mathnotation{\Entr}{S}				
\mathnotation{\Ener}{E}				
\mathnotation{\IEner}{U}			
\mathnotation{\Pres}{P}				
\mathnotation{\Prest}{{\mathbb{\Pres}}}		
\mathnotation{\pres}{p}				
\mathnotation{\unitI}{{\mathbb{I}}}		
\mathnotation{\Q}{Q}				
\mathnotation{\Qv}{{\mathbf{\Q}}}		
\mathnotation{\gasc}{R}				
\mathnotation{\Afp}{A}				
\mathnotation{\Bfp}{B}				
\mathnotation{\colfreq}{\nu}			
\mathnotation{\fdist}{f}			
\mathnotation{\fdistt}{\hat{f}}		
\mathnotation{\inv}{\psi}			
\mathnotation{\velT}{\varpi}			
\mathnotation{\Pra}{\sigma}			
\mathnotation{\visc}{\mu}			
\mathnotation{\kvisc}{\nu}			
\mathnotation{\tcond}{\eta}			
\mathnotation{\tdiff}{\kappa}			
\mathnotation{\ros}{E}				
\mathnotation{\rost}{\mathbb{\ros}}		
\mathnotation{\tros}{\Trless{\ros}}		
\mathnotation{\trost}{\Trless{\rost}}		
\mathnotation{\vort}{\Omega}			
\mathnotation{\vortv}{\mathbf{\Omega}}		
\mathnotation{\Cv}{C_v}				
\mathnotation{\Cp}{C_p}				
\mathnotation{\DtS}{\Sigma}
\mathnotation{\divPp}{\Psi}
\mathnotation{\divPpv}{\boldsymbol{\divPp}}
\mathnotation{\pdegree}{d}			
\mathnotation{\pcoeff}{\alpha}			
\mathnotation{\ccz}{\pcoeff^{(0)}}
\mathnotation{\cci}{\pcoeff^{(1)}}
\mathnotation{\ccii}{\pcoeff^{(2)}}
\mathnotation{\cciii}{\pcoeff^{(3)}}
\mathnotation{\cctz}{\tilde\pcoeff^{(0)}}
\mathnotation{\ccti}{\tilde\pcoeff^{(1)}}
\mathnotation{\cctii}{\tilde\pcoeff^{(2)}}
\mathnotation{\cctiii}{\tilde\pcoeff^{(3)}}
\mathnotation{\cctiv}{\tilde\pcoeff^{(4)}}
\mathnotation{\cctv}{\tilde\pcoeff^{(5)}}
\mathnotation{\cctvi}{\tilde\pcoeff^{(6)}}
\mathnotation{\Tens}{A}				
\mathnotation{\Tenst}{{\mathbb{\Tens}}}		
\mathnotation{\wvec}{w}				
\mathnotation{\wvecv}{{\mathbf{\wvec}}}		

\newcommand{\CE}{Chapman--Enskog}
\newcommand{\NS}{Navier--Stokes}
\newcommand{\MB}{Maxwell--Boltzmann}

\begin{document}

\title{Higher-order Continuum Approximation for Rarefied Gases}
\author{Edward A. Spiegel}
\email{eas@astro.columbia.edu}
\affiliation{Department of Astronomy, Columbia University, New York, NY 10027}
\author{Jean-Luc Thiffeault}
\email{jeanluc@mailaps.org}
\affiliation{Department of Applied Physics and Applied
Mathematics, Columbia University, New York, NY 10027}
\altaffiliation[Present address: ]{Department of Mathematics, Imperial College
  London, SW7 2AZ, United Kingdom.}

\begin{abstract}
The Hilbert--\CE\ expansion of the kinetic equations in mean flight times is
believed to be asymptotic rather than convergent.  It is therefore inadvisable
to use lower order results to simplify the current approximation as is done in
the traditional \CE\ procedure, since that is an iterative method.  By
avoiding such recycling of lower order results, one obtains macroscopic
equations that are asymptotically equivalent to the ones found in the \CE\
approach.  The new equations contain higher order terms that are discarded in
the \CE\ method.  These make a significant impact on the results for such
problems as ultrasound propagation.  In this paper, it is shown that these
results turn out well with relatively little complication when the expansions
are carried to second order in the mean free time, for the example of the
relaxation or BGK model of kinetic theory.
\end{abstract}


\maketitle

\section{Introduction}
\label{sec:intro} 

Derivations of the dynamical equations for a continuum under the influence of
internal friction were first produced in the nineteenth century by
phenomenological arguments in the works of Navier, Poisson and Stokes.  These
``\NS'' equations contained coefficients of viscosity and heat conduction
whose evaluation was not provided by the early phenomenology. It was not until
Maxwell's treatment of the problem through elementary kinetic theory that the
dependences of these coefficients on the thermodynamical state variables began
to be understood.  When the need for a kinetic theoretical underpinning of the
theory of the dissipative properties of fluids had become clearer,
Hilbert~\cite{Hilbert1912} developed approximate solutions of Boltzmann's
basic kinetic equation in terms of an expansion in the mean flight time of the
particles constituting the fluid.  Finally, in the hands of Chapman and of
Enskog, the derivation of the equations of fluid dynamics from the basic
kinetic theory of gases took on the form that was most widely used throughout
the last century.  This theory, in the formal version that Enskog developed,
was summarized authoritatively in the book of Chapman and
Cowling~\cite{ChapmanCowling2nd}, as well as by a number of later
books~\cite{Uhlenbeck,CercignaniBoltz2,FerzigerKaper,Kennard}.

Though it was satisfying that the same equations came from phenomenology as
from kinetic theory, there was also bad news: when the mean flight time is
long, the descriptions of sound propagation and of shock wave structure from
the \NS\ equations are not very good when compared to
experiment~\cite{Uhlenbeck}.  This is one of the reasons that other approaches
to the derivation of fluid dynamical theory have been sought over the years.
The first approach was naturally to extend the Hilbert development to higher
order.  This was achieved by Burnett~\cite{Burnett1935}, who went to second
and third order in the expansions.  However the form of Burnett's equations
that are conventionally employed are those that result when the Euler
equations have been introduced into them as Chapman and
Cowling~\cite{ChapmanCowling2nd} did (see the discussion by Agarwal
\etal~\cite{Agarwal2001}).  This may be the reason for the disappointing
result that little or no improvement was found (and with often unphysical
consequences) when the reduced Burnett equations were used for situations when
the mean flight time was not extremely small.  The reason that going to higher
order does not lead to any real improvement in the results obtained with the
\CE\ approach is probably that, as Grad~\cite{Grad1963} and Uhlenbeck and
Ford~\cite{Uhlenbeck} have suggested, the Hilbert expansion is asymptotic
rather than convergent.  Since the \CE\ procedure is a sort of iteration
process, it would seem to be inappropriate for a series that is not
convergent.  That is, in the \CE\ procedure, one uses results from the
previous order to simplify the results in the current order and this has the
effect of restricting the domain of validity of the approximation at any
order.

To circumvent this limitation of the \CE\ procedure, an alternate approach was
suggested in which the recycling of lower order results was
avoided~\cite{Chen2000,ChenThesis,Chen2001a,Chen2002}.  In the case of the
relaxation model of kinetic theory~\cite{Bhatnagar1954,Welander1954} this
extension of the procedures was straightforward to apply and the results were
encouraging.  (The alternate approach was also carried out to first order for
a Fokker--Planck type collision operator in
Ref.~\cite{SpiegelThiffeault2003}.)  Not only was an improved description of
the structures of shock waves over that from the \NS\ equations obtained, but
also the phase speed of sound waves that emerged was in good agreement with
experiment for all ratios of mean flight times to wave periods, unlike the
\NS\ results.  Though the results for the damping of sound waves did not turn
out as well, they were not bad for mean flight times below the sound period
and they were better than the \NS\ results.  At present though, it is unclear
whether a continuum theory can do much better than this result for damping.
Perhaps a microscopic theory is required to do better, as
Stubbe~\cite{Stubbe1990} and Sukhorukov and
Stubbe~\cite{Sukhorukov1995,Stubbe1999} suggest.  Here we further probe the
continuum theory by going to higher order using the new approach.

\section{Basics}

We consider a gas of $N$ identical particles of mass $m$ whose individual
positions and velocities are denoted by $\xv$ and $\vv$.  The
distribution of the particles in the phase space with coordinates $x_i$ and
$v_j$ ($i,j=1,2,3$) is represented by a swarm of representative points whose
local density $f(\xv, \vv, t)$ is defined such that the probable
number of particles in the volume element $d\xv\, d \vv$ is
$f(\xv, \vv, t)\, d\xv\, d\vv$.

The fluid dynamical variables are defined in terms of moments of $f$.  The
density and fluid velocity are
\begin{equation}
	\dens \ldef \int \mass\fdist\, d\vv \qquad {\rm and}\qquad
	\uv \ldef \frac{1}{\dens}\int \mass\vv\,\fdist d\vv
\end{equation}
and, with the peculiar velocity $\cv \ldef \vv - \uv$, we may define the
temperature
\begin{equation}
	\Temp \ldef \frac{\mass}{3\gasc\,\dens} \int \cc^2\fdist\dint^3\vc\, .
	\label{eq:temp}
\end{equation} 
We seek equations for these quantities.  Interest has been focused on them
because, as van Kampen~\cite{VanKampen1987} has stressed, these are the
macroscopic ``slow variables'' that emerge from Boltzmann's theory.

The particles obey Hamiltonian dynamics and so the phase fluid flows
incompressibly and satisfies an equation of the form
\begin{equation}
  \Liouv f = \alpha - \beta f
  \label{kineq}
\end{equation}
where
\begin{equation}
  \Liouv \ldef  \partial_t  + v^i \partial_i + a^j \partial_{v^j}\,\,
\end{equation}
and summation over repeated symbols is understood throughout this work.  The
quantity $\accelv$ is the imposed external force per unit mass.

On the left of~\eqref{kineq}, we have the rate of change of $f$ resulting from
the streaming of the representative points through phase space.  On the right,
we see the effects of the interactions of the particles with those in their
surroundings.  Particles with velocity equal to $\vv$ have their velocities
changed to some other values at a rate $\beta f$ per unit volume while
particles with other velocites are given the velocity $\vv$ at the rate
$\alpha$.  In strict equilibrium, $f$ has the value $f_0$, which does not
depend on $\xv$ or $t$.  Then the left side of~\eqref{kineq} vanishes and we
find that
\begin{equation}  
  f_0 = \alpha/\beta.
\label{KP} \end{equation}
This expression recalls the Kirchhoff--Planck law of radiation theory and
hints that radiative transfer theory is one source of inspiration for the
relaxation model~\cite{Spiegel1972}.

According to the description we adopt of the way the particles interact with
each other, we may find $f_0$ either as a consequence of the vanishing of the
interaction term or we may need to prescribe it.  In either case, the choice
of greatest interest is the \MB\ distribution
\begin{equation}
	\fdist_0(\xv,\vv,\time) = \frac{\ndens}{(2\pi\gasc\Temp)^{3/2}}\,
		\exp\l(-\frac{\cc^2}{2\gasc\,\Temp}\r),
	\label{eq:Maxwellian}
\end{equation}
where $n\ldef\rho/m$.  In the case where the fluid variables are allowed to
depend on space and time, the use of the \MB\ distribution for $f_0$ defines a
local thermodynamic equilbirium.  When we
introduce~\eqref{KP}--\eqref{eq:Maxwellian} into the kinetic equation, with the
understanding that $f_0$ represents only a local equilibrium, we get the
relaxation model
\begin{equation}
  \Liouv  f = \frac{f_0 - f}{\relt}
  \label{eq:kinetic}
\end{equation}
where $\relt\ldef 1/\beta$.  We may then study the case of nonequilibrium,
with $f$ and $f_0$ differing, by using the kinetic equation to follow the
evolution of $f$.  Alternatively, we may seek the coarser description of fluid
dynamics in terms of the moments of the kinetic equation.

Given the definitions of the fluid variables and the expression of $f_0$ in
terms of them, we obtain consistency between the two by imposing the 
matching conditions~\cite[pp.~362 and~369]{ClemmowDougherty},
\begin{equation}
  \int \left(f_0 - f\right) \psi^\mu \dint^3\vc = 0
  \label{match}
\end{equation}
where
\begin{equation}
  \psi^\mu = m\l(1\,,\, v^i\,,\, \tfrac{1}{2}c^2\r)
\end{equation}
with $\mu=0,1,2,3,4$ and $i=1,2,3$.  This ensures that the state variables in
$f_0$ are the same ones that are given by the moments of $f$.  Since we shall
assume here that $\tau$ does not depend explicitly on $\mathbf{v}$, though it
may depend on the state variables, the matching conditions are equivalent to
the conservation of mass, momentum and energy.

If we now multiply the kinetic
equation by $\psi^\mu$ and integrate over velocity space, we obtain the fluid
equations
\begin{align}
\pdt\dens + \div(\dens\,\uv) &= 0,
	\label{eq:denseq}\\
\pdt\uv + \uv\cdot\grad\uv &= -\dens^{-1}\div\Prest
		+ \accelv,
	\label{eq:ueq}\\
\threehalf\gasc\dens\l(\pdt\Temp + \uv\cdot\grad\Temp\r)
		&= -\Prest:\grad\uv - \div\Qv
	\label{eq:Tempeq}
\end{align}
where
\begin{equation}
	\Prest \ldef \int\mass\cv\cv\fdist\dint^3\vc \qquad {\rm and} \qquad
	\Qv \ldef \int\half\mass\cc^2\cv\fdist\dint^3\vc
	\label{eq:PrestQvdef}
\end{equation}
are the pressure tensor and heat flux vector.  The trace of the
pressure tensor is $3\pres$ where
\begin{equation}
  \pres \ldef \gasc \rho\, \Temp\, ;
  \label{state}
\end{equation}  
here,  $\gasc=k/m$ is the gas constant and $k$ is Boltzmann's constant. 

Another useful quantity is
 \begin{equation}
  S \ldef \Cv\,\ln \l( \frac{p}{\rho^\gamma} \r)
  \label{ent}
\end{equation}
where $\Cv$ is the specific heat at constant volume and $\gamma$ is the ratio
of specific heats; $\gamma=5/3$ in this work and $\Cv=\frac{3}{2}\gasc$.  The
expression for $S$ is that for the specific entropy of a perfect gas when
$\pres$ is the pressure.  On using the continuity equation and the equation of
state, we see that
\begin{equation}
  (\partial_t + \uv \cdot \grad)
  \l(\frac{\Entr}{\Cv}\r) = \Dt{\ln{\Temp}} + \tfrac{2}{3}\,\div\uv.
\end{equation}
We may then rewrite the heat equation~\eqref{eq:Tempeq} as
\begin{equation}
\dens T \l( \pdt S + \uv\cdot\grad S \r)
		= - \left( \Prest - p\,\mathbb{I} \right) :\grad\uv - \div\Qv
	\label{eq:Sequat}
\end{equation}
where $\unitI$ is the unit tensor.  This result motivates the introduction of
a notation for the tensors of kinetic theory, namely that an $\times$
over any tensor designates its trace while a circle over a tensor, say
$\Tenst$, signifies its symmetrized and traceless part, as in
\begin{equation}
	\Trless{\Tenst} \ldef \half\,(\Tenst + \transp{\Tenst})
		- \third\,\Tr{\Tenst}\,\unitI \, ,
		\label{eq:trlessdef}
\end{equation}
where the overtilde indicates a transpose.  For example, we see from the
definition of the stress tensor that $\Tr{\mathbb{P}}=3p$ and
$\Trless{\mathbb{P}}= 0$.

To obtain
determinate expressions for $\Prest$ and $\Qv$, we seek an approximation for
$f$.  In what follows we describe  a procedure for obtaining and using one.
We begin with a standard  expansion of $f$ in terms of $\relt$:
\begin{equation}
  f = f_0 + f_1\, \relt + f_2\, \relt^2 + \ldots.
  \label{expansion}
\end{equation}
When we introduce $f_n\relt^n$ (no summation) into the formulae for $\Prest$
and $\Qv$ we obtain contributions written as $\Prest^{(n)}$ and $\Qv^{(n)}$
which lead to
\begin{equation}
  \Prest = \Prest^{(0)} +  \Prest^{(1)} + \ldots \qquad \text{and} \qquad 
  \Qv = \Qv^{(0)} + \Qv^{(1)} + \ldots.
  \label{PNQN}
\end{equation}

At the lowest order, in
which $f_0$ is the \MB\  distribution, we find that 
$\Prest^{(0)}=\pres\,\unitI$ and
$\Qv^{(0)}=\mathbf{0}$, and so we have~\eqref{eq:denseq} and
\begin{align}
\pdt\uv + \uv\cdot\grad\uv +\dens^{-1}\grad p -
		\accelv &=\Order{\tau},
	\label{eq:ueuler}\\
\threehalf\gasc\dens\l(\pdt\Temp + \uv\cdot\grad\Temp\r)
		+ p \div\uv &=  \Order{\tau}.
	\label{eq:Tempeq0}
\end{align}
Without the $\Order{\tau}$ terms, these are the Euler equations.

At the next order, we obtain the fluid equations found earlier
\cite{Chen2000,Chen2001a}; these reduce to the standard \NS\ equations when
$\relt$ becomes strictly infinitesimal.  As these first-order developments are
central to the current work we present them next
(Section~\ref{sec:firstorder}).  This provides us with an opportunity to
describe how the \NS\ equations emerge and to indicate how the Lorentz force
enters the theory when electromagnetic effects are considered.  Then we
proceed to carry the development on to the next order
(Section~\ref{sec:secondorder}) and to show the predictions of the
second-order theory for the propagation of free sound waves
(Section~\ref{sec:ultrasonic}).

\section{First-order Theory}
\label{sec:firstorder}

\subsection{Approximation for $f$}

At order~$\relt$, we readily obtain from Eq.~\eqref{eq:kinetic} that
\begin{equation}
	\fdist_1 = -\,\Liouv\fdist_0\, .
	\label{eq:fdist1eq}
\end{equation}
It proves advantageous to multiply~\eqref{eq:fdist1eq} 
by $1/\fdist_0$ and to rewrite it as
\begin{equation}
	\fdistt_1 = -\Liouv\ln\fdist_0,
	\label{eq:fdistt1eq}
\end{equation}
where~\hbox{$\fdistt_1 \ldef \fdist_1/\fdist_0$}.
Then, with \eqref{eq:Maxwellian}, we have
\begin{equation}
	\fdistt_1 = -	\Liouv \l[ -\frac{\cc^2}{2\gasc\,\Temp}
		+ \ln\frac{\ndens}{(2\pi\gasc\Temp)^{3/2}} \r] ,
\end{equation}
which we may write out as
\begin{equation}
\fdistt_1 = 		
	\frac{1}{\gasc\,\Temp}\,\cv\cdot\Liouv\cv
		- \l( \frac{\cc^2}{2\gasc\,\Temp} - \frac{3}{2} \r)
			\Liouv\ln\Temp
		- \Liouv\ln{\dens}.
	\label{eq:RHSorder1}
\end{equation}

\subsection{Developments}

Some further rearrangements are usefully made on noticing that
\begin{equation}
	\Liouv\cv = \Liouv(\vv - \uv)
	= \accelv - \Dt{\uv} - (\cv\cdot\grad)\uv
	\label{eq:pecvelderiv}
\end{equation}
where $D_t = \partial_t + \uv \cdot \grad$.
We can then rewrite  Eq.~\eqref{eq:RHSorder1} as
\begin{multline}
\fdistt_1 = \frac{1}{\gasc\,\Temp}\,\cv\cdot \l[ \accelv - \Dt{\uv}
		- (\cv\cdot\grad)\uv \r] \\
		- \l(  \frac{\cc^2}{2\gasc\,\Temp} - \frac{3}{2}  \r)
			\l(\Dt{\ln{\Temp}} + \cv\cdot\grad\ln{\Temp}\r)
		- \Dt{\ln{\dens}} - \cv\cdot\grad\ln{\dens}\,.
	\label{eq:Liouvf0}
\end{multline}

It is useful at this point to introduce the rate-of-strain tensor
\begin{equation}
  \rost \ldef \half(\grad\uv + \transp{\grad\uv})
  \label{eq:rostrain}
\end{equation}
where the tilde over a tensor indicates transposition.
Then, the trace $\Tr{\rost} = \div{\mathbf{u}}$ and we
may rewrite the~$\cv\cdot(\cv\cdot\grad)\uv$ term 
on the right of~\eqref{eq:Liouvf0} as
\begin{equation}
  \cv\cdot(\cv\cdot\grad)\uv =
\cv\, \cv:\Trless{\rost} + \frac{1}{3} c^2 \grad \cdot \mathbf{u} \, .
\end{equation}

After introducing the definition of $S$ and using the ideal gas
law~\eqref{state} and the continuity equation~\eqref{eq:denseq}, we find by a
longish series of straightforward substitutions that~\eqref{eq:Liouvf0}
becomes
\begin{equation}
\fdistt_1 = 
	-\l(\frac{\cc^2}{2\gasc\,\Temp} - \frac{5}{2}\r)
				\cv\cdot\grad\ln{\Temp}
		- \frac{1}{\gasc\,\Temp}\,\cv\,\cv:\trost
		+ \cv\cdot\divPpv
		- \l(\frac{\cc^2}{2\gasc\,\Temp} - \frac{3}{2}\r)\DtS
	\label{eq:rhsint2}
\end{equation}
where
\begin{equation}
  \Sigma \ldef D_t  \l(  \frac{S}{\Cv}  \r)
  \label{eq:DtSdef}
\end{equation}
and
\begin{equation}
  \divPpv \ldef -\frac{1}{\gasc\,\Temp}\, \l( \Dt{\uv} - \accelv \r)
  - \grad\ln\pres \,.
  \label{eq:divPpdef}
\end{equation}
On using~\eqref{eq:ueq},  we may also write
\begin{equation}
 \divPpv =  \frac{1}{\pres}\,\div\Trless{\Prest}  
 \label{eq:divPp2}
\end{equation}

In a final cosmetic  touch, we  define the dimensionless peculiar velocity
\begin{equation}
	\cnuv \ldef \mathbf{c}/  \sqrt{2\gasc\,\Temp}
\end{equation}
and rewrite~\eqref{eq:rhsint2} as
 \begin{equation}
\fdistt_1 = 
	- \l(  \cnu^2 - \frac{5}{2}  \r)  \cv\cdot\grad\ln{\Temp}
		- 2\,\cnuv\,\cnuv:\trost
		+ \,\cv\cdot\divPpv
		- \l( \cnu^2 - \frac{3}{2} \r) \DtS\,.
		\label{eq:rhsint3}
\end{equation}

\subsection{The Prandtl Number}
\label{sec:prandtl}

The relaxation model is prone to some ambiguity coming from the the lack of
specificity in the details of the interactions of particles.  In particular,
the model leaves open the way $\tau$ depends on $\mathbf{v}$.  This is really
a matter that is related to the nature of the atomic interactions and does not
seem to have a significant qualitative influence on the macroscopic results.
However, the Prandtl number, $\Pra$, given by the relaxation model is
different from that predicted by the Boltzmann model which is restricted to
two-body interactions.  (The Prandtl number is the ratio of the viscosity
times $\Cp$ to the thermal conductivity.)  Experiments in noble gases give
values for $\Pra$ close to the value $2/3$ predicted by the Boltzmann equation
for a gas of hard spheres; by contrast, the value unity is found with the
relaxation model.  In calculating macroscopic quantities with the relaxation
model, notably the stress tensor and energy flux, it is sometimes considered
desirable to adjust the value of $\Pra$ in the results to be $2/3$.  Since
there is more than one way to do this, a systematic procedure seems called
for, particularly when the Lorentz force comes into play.

Woods~\cite{Woods} has given a prescription for introducing the desired value
of the Prandtl number that seems to work well, though it is more of a rule of
thumb than a fully justified procedure.  It has the advantage of providing a
definite prescription that can be carried into higher order and it leads to
agreement with results from the Boltzmann equation for \emph{all} of the
numerical coefficients that appear in the higher moments obtained from the
\CE\ procedure in second order in $\tau$.  Since this modification of the
results can be retracted by setting $\sigma=1$, it is harmless and we shall
introduce it.

In Woods' approach one introduces the Prandtl number into the formulae for
approximating $f$ rather than into the final fluid equations.  The motivation
for Woods' procedure is that the times of relaxation for momentum and energy
are different \cite{ChandraStellarDynamics} so that he has suggested the use
of different relaxation times depending on whether a term from $f$ enters the
pressure tensor or the heat flux.  Terms with even powers in~$\cv$ contribute
to the pressure tensor and are assigned a relaxation time~$\relt$; terms with
odd powers in~$\cv$ contribute to the heat flux and are assigned a relaxation
time~$\relt/\Pra$ where $\Pra$ is the Prandtl number.  Therefore, we
rewrite~\eqref{eq:rhsint3} as
\begin{equation}
	\fdistt_1 = -\frac{1}{\Pra}
		\l(\cnu^2 - \frac{5}{2}\r)\cv\cdot\grad\ln{\Temp}
		- 2\cnuv\,\cnuv:\trost
		+ \frac{1}{\Pra}\,\cv\cdot\divPpv
		- \l( \cnu^2 - \frac{3}{2} \r) \DtS\,.
	\label{eq:fdistt1}
\end{equation}
This way of adjusting the Prandtl number differs from that in the first-order
derivation of Refs.~\cite{Chen2000,Chen2001a} and, for $\Pra=2/3$, it leads to
agreement of our second-order theory with Burnett's results~\cite{Burnett1935}
from the Boltzmann equation if we discard terms of order $\relt^3$ (see
the appendix).

If the acceleration~$\accelv$ is not a function of~$\vv$,
Eq.~\eqref{eq:fdistt1} is cubic in~$\cc^i$, so~$\fdistt_1$ is a polynomial
in~$\cc^i$.  If~$\accelv$ is a polynomial in~$\vv$, then we can still
write~$\fdistt_1$ as a polynomial in~$\cv$, possibly of higher order than
cubic.  An interesting instance of this arises when $\accelv$ includes the
Lorentz force on charged particles.  The only term in \eqref{eq:fdistt1} that
involves $\accelv$ is $\divPpv$.  That term vanishes for the standard \CE\
expansion without the electric field $\Ev$ and magnetic field $\Bv$.  When
$\Ev$ and $\Bv$ are included, $\divPpv$ is nonzero and can depend on velocity
through~$\accelv$ [Eq.~\eqref{eq:divPpdef}].  Let
\begin{equation}
	\accelv(\xv,\vv,\time)
		= \frac{\elc}{\mass}\l[
			\Ev(\xv,\time)
			+ \vv\times\Bv(\xv,\time)\r]
\end{equation}
be the acceleration due to the Lorentz force, where $\elc$ is the electron
charge in MKS units.  By rewriting $\accelv$ as
\begin{equation}
	\accelv(\xv,\vv,\time)
		= \frac{\elc}{\mass}\l[
			\Ev(\xv,\time)
			+ \uv\times\Bv(\xv,\time)\r]
		+ \frac{e}{\mass}\,\cv\times\Bv(\xv,\time)\,.
\end{equation}
we split the acceleration into a part that is independent of $\cv$ and a part
linear in~$\cv$.  As we see from Eq.~\eqref{eq:fdistt1},~$\accelv$ will be
dotted with~$\cv$, and so $\cv \times \Bv$ will not contribute.  Hence,
electric and magnetic fields can be included by simply expressing the
acceleration as
\begin{equation}
	\accelv(\xv,\time) = \frac{\elc}{\mass}\l[
		\Ev(\xv,\time) + \uv\times\Bv(\xv,\time)\r],
\end{equation}
which is independent of~$\vv$.

\subsection{Matching Conditions}
\label{sec:matching1}

Although the relaxation model is often regarded as linear, in much the same
way that the analogous equation of radiative transfer is, it has to be
recognized that $f_0$ depends on the moments of the full $f$ and so cannot be
considered as given $\it{a\ priori}$.  Treating the problem as formally linear
is possible because the feedback on the fluid variables from alterations in
$f$ are forced to be consistent with the basic equations by the matching
conditions~\eqref{match}.  In the present work, they guarantee the
conservation of the collisional invariants $\psi^\mu$.

If we insert the solution~\eqref{eq:fdistt1} for~$\fdist_1$
into~\eqref{match} we find that the density matching condition is
satisfied to this order.  The momentum matching condition is that 
\begin{equation}
	\int   \mass\vv\,\fdist_1 \dint^3\vc =
	\frac{\dens \gasc\,\Temp}{2\Pra}\, \divPpv =
		\frac{1}{2\Pra}\,\div(\Prest - \pres\,\unitI).
	\label{eq:1stordermommatch}
\end{equation}
But $\mathbb{P}^{(0)}=\pres\,\mathbb{I}$ so that $\mathbb{P}-\pres\,\mathbb{I}
= \Order{\tau}$.  Hence, the momentum matching condition is satisfied to the
order of the current approximation, which is all that can be asked in an
asymptotic development.

Similarly, the energy matching condition is that
\begin{equation}
	\tfrac{1}{2}   \int   \mass\vc^2\,\fdist_1 \dint^3\vc =
	-\tfrac{3}{4}\dens \gasc\Temp\, \DtS.
	\label{eq:1storderenermatch}
\end{equation}
with $\Sigma$ given by~\eqref{eq:DtSdef}.
From~\eqref{eq:divPpdef} and~\eqref{eq:Tempeq}, 
we find that~\eqref{eq:1storderenermatch} is
\begin{equation}
	\DtS = \Dt{\ln{\Temp}} + \tfrac{2}{3}\,\div\uv
	= -\frac{2}{3\pres}  \Trless{\Prest}:\grad\uv + \div\Qv\,.
	\label{eq:DtS2}
\end{equation}
On recalling that $\Qv^{(0)}=0$, we see that $\Sigma$ is~$\Order{\relt}$.
We thus conclude that all the matching conditions are satisfied to the
required order.

\subsection{Pressure Tensor and Heat Flux}
\label{sec:prestheatf1}

As mentioned in Section~\ref{sec:firstorder}, if the acceleration~$\accelv$
does not depend on~$\vv$ (except possibly for a magnetic term) and~$\relt$
does not depend on~$\vv$, then we can write~$\fdistt_1$ as a third degree
polynomial in the components of~$\vv$ .  Finding the contributions of the
first-order terms in $f$ to the pressure tensor and heat flux is then just a
matter of inserting the solution~\eqref{eq:fdistt1} for~$\fdistt_1$ (times
$\tau$) into~\eqref{eq:PrestQvdef} and carrying out the resulting integrals,
as indicated by~\eqref{PNQN}.  We find
\begin{equation} 
	\Prest^{(1)} = -\pres\,\relt\bigl(2\trost + \DtS\,\unitI\bigr),
	\qquad
	\Qv^{(1)} = -\tfrac{5}{2}\,\frac{\relt}{\Pra}\,\pres\,\gasc\,\Temp\l(
		\grad\ln{\Temp} - \divPpv\r).
	\label{eq:PrestQv1}
\end{equation}
Here we see the appearance of the viscosity and the conductivity,
\begin{equation}
  \visc \ldef \pres\,\relt \qquad
  \text{and} \qquad
  \tcond \ldef \frac{1}{\sigma}\, \Cp \,\pres\, \relt
  \label{eq:diff}
\end{equation}
where we have written $\Cp$ for $5\gasc/2$.  The Prandtl number is
$\sigma = \visc\, \Cp/ \tcond$.

The first terms in each of the expressions (\ref{eq:PrestQv1}) are those that
appear in \NS\ equations.  The second terms (proportional to~$\DtS$
and~$\divPpv$) are the terms that are retained when the additional
approximations introduced in the usual \CE\ procedure are not made.  That is,
if we had used the zeroth-order results (the Euler equations) to eliminate the
time derivatives contained in~$\DtS$ and~$\divPpv$, we would have obtained the
\NS\ equations, as in the \CE\ method, though generally speaking this extra
(and unnecessary) step is normally taken earlier in the development in the
standard approach.  Our suggestion is that if one wants to simplify the
expressions for $\Prest^{(1)}$ and $\Qv^{(1)}$ by eliminating the time
derivatives, it is more accurate to use the current approximation for the
equations of motion rather than a lower-order one.  If this elimination is
done iteratively, it shows that our expressions for the higher moments contain
terms of all orders in $\tau$.  However, we have left the time derivatives as
they appear in~$\DtS$ and~$\divPp$ [Eqs.~\eqref{eq:DtSdef}
and~\eqref{eq:divPpdef}], and we shall simply work with the full system
self-consistently when we illustrate a physical problem in
Section~\ref{sec:ultrasonic}.
 
The stress tensor and energy flux
at this stage are written as
\begin{equation}  \Prest =  \Prest^{(0)} + \Prest^{(1)}
  + \Order{\relt^2} \qquad \textrm{and} 
  \qquad \Qv = \Qv^{(0)} + \Qv^{(1)} + \Order{\relt^2}
\end{equation}
where $\Prest^{(0)}=p\,\unitI$ and $\Qv^{(0)}=\mathbf{0}$.  Since the terms by
which $\Prest^{(1)}$ and $\Qv^{(1)}$ differ from the corresponding \NS\
expressions are $\Order{\relt^2}$, the differences are within the allowed
error estimates for both treatments.  By the same token, we find directly from
the definition of $\Prest$ that its trace is $3p$ while the trace of our
expression for it differs from $3p$ by terms of order $\tau^2$.  This is also
compatible with the asymptotics to this order.  As we shall see, in the next
order, the trace of our approximation to $\Prest$ differs from $3p$ only by
terms of order $\tau^3$.

A suggestive way look at these results is to observe that we may write
\begin{equation}
  \Prest = \left(p-\relt D_t\, p\right)\unitI
  - p \relt \left(\gamma \div\uv \unitI
  - 2\trost \right) +  \Order{\relt^2}.
  \label{otherway}
\end{equation}
To this approximation, if we replace $\pres$ by the first two terms in its
Taylor series in time, we have
\begin{equation}
  \Prest = \pres(t-\relt)\unitI
  - \mu \left(\gamma\div\uv\, \unitI - 2\trost \right) +
  \Order{\relt^2}.
  \label{delay}
\end{equation}
That is, the form is like that of the \NS\ equations except that 
the pressure is evaluated at one relaxation time prior to the
present, and there is a volume (or bulk) viscosity term. 
 
\section{Second-order Theory}
\label{sec:secondorder}

\subsection{The Distribution Function}
\label{sec:distr-funct2}
In second order in~$\relt$, the kinetic equation~\eqref{eq:kinetic} yields
\begin{equation}
	\tau\fdist_2 = - \Liouv(\tau\fdist_1)\, ,
	\label{eq:fdist2eq}
\end{equation}
which in terms of~\hbox{$\fdistt_2\ldef\fdist_2/\fdist_0$} becomes
\begin{equation}
	\fdistt_2 = - \Liouv\fdistt_1
	-  \fdistt_1\,\Liouv\ln\fdist_0\, - \fdistt_1 \Liouv \ln\tau\, .
	\label{eq:fdistt2eq}
\end{equation}

We allow for a dependence of $\tau$ on the macroscopic state variables, $p$
and $T$, but assume that $\tau$ does not depend on the particle velocities,
either microscopic or macroscopic.   We then write
\begin{equation}
  \Liouv \ln \tau = r D_t\ln \rho + s D_t \ln T
\end{equation}
where
\begin{equation}\label{xizeta}
r = \frac{\partial \ln \tau}{\partial \ln \rho}
\qquad \textrm{and} \qquad
s = \frac{\partial \ln \tau}{\partial \ln T}\,.
\end{equation}
Chapman and Cowling specify $r=-1$ on the authority of Maxwell 
while,  depending on the density of the gas,~$\sT$ varies from
about~$0.5$ to~$1$~\cite[p.~223]{ChapmanCowling2nd}.

With $r=-1$, the third term on the right-hand side
of~\eqref{eq:fdistt2eq} is 
\begin{equation}
\fdistt_1\,\Liouv\ln\relt
	= \fdistt_1\,\l(\sT\,\Liouv\ln\Temp - \Liouv\ln\pres\r).
\end{equation}
After introduction of the fluid
equations~\eqref{eq:denseq}--\eqref{eq:Tempeq}, this becomes
\begin{equation}
\fdistt_1\,\Liouv\ln\relt
	= \fdistt_1\,\l[\tfrac{1}{3}(5 - 2\sT)\div\uv
		+ (\sT - 1)\DtS
		+ \sT\,\cv\cdot\grad\ln\Temp - \cv\cdot\grad\ln\pres\r].
	\label{eq:Liouvrelt}
\end{equation}

The middle term on the right of Eq.~\eqref{eq:fdistt2eq} may be written
out as
\begin{multline}
	\fdistt_1\,\Liouv\ln\fdist_0
	= -\l[\frac{1}{\Pra}
		\l(\cnu^2 - \frac{5}{2}\r)\cv\cdot\grad\ln{\Temp}
		+ 2\cnuv\,\cnuv:\trost
		- \frac{1}{\Pra}\,\cv\cdot\divPpv
		+ \l(\cnu^2 - \frac{3}{2}\r)\DtS\r]\\
\times\l[\l(\cnu^2 - \frac{5}{2}\r)\cv\cdot\grad\ln{\Temp}
		+ 2\cnuv\,\cnuv:\trost
		- \cv\cdot\divPpv + \l(\cnu^2 - \frac{3}{2}\r)\DtS\r].
	\label{eq:quadterm}
\end{multline}

The first of the three terms on the right of Eq.~\eqref{eq:fdistt2eq} requires
a certain attentiveness so it is well to divide its calculation into
manageable pieces.  Firstly, the derivative of the peculiar
velocity~\eqref{eq:pecvelderiv} can be written as
\begin{equation}
	\Liouv\cv =  \RTv^2 (\divPpv + \grad\ln\pres)
		- \cv\cdot\rost + \cv\times\vortv,
	\label{eq:pecvelderiv2}
\end{equation}
where
\begin{equation}
  \RTv^2 \ldef 2\gasc\Temp,
  \label{eq:RTVdef}
\end{equation}
and~\hbox{$\vortv \ldef \half\curl\uv$} is the vorticity.
When we observe that
\begin{equation}
	\Liouv \RTv = \half \RTv
	\,\l(\DtS - \tfrac{2}{3}\div\uv + \cv\cdot\grad\ln\Temp\r);
\end{equation} we find that~\eqref{eq:pecvelderiv2} implies
\begin{equation}
	\Liouv\cnuv = \Liouv\l(\frac{\cv}{2RT}\r)
	= \RTv(\divPpv + \grad\ln\pres)
		- \cnuv\cdot\trost + \cnuv\times\vortv
		- \half\,\cnuv(\DtS + \cv\cdot\grad\ln\Temp).
\end{equation}
Because it is antisymmetric, the vorticity term
disappears from~$\Liouv\cc^2$:
\begin{equation}
	\Liouv\cc^2 = 2\cv\cdot\Liouv\cv
	= \RTv^2(\cv\cdot\divPpv + \cv\cdot\grad\ln\pres)
		- 2\cv\,\cv:\rost\,,
\end{equation}
and similarly 
\begin{equation}
	\Liouv\cnu^2 = \RTv\,(\cnuv\cdot\divPpv + \cnuv\cdot\grad\ln\pres)
		- 2\cnuv\,\cnuv:\trost
		- \cnu^2(\DtS + \cv\cdot\grad\ln\Temp)\,.
\end{equation}

Next we have
\begin{align}
	\Temp\,\Liouv\grad\ln\Temp &=
	\Dt{\grad\Temp} + \cv\cdot\grad\grad\Temp
		+ \tfrac{2}{3}\,\div\uv\,\grad\Temp
		- \grad\ln\Temp\,(\cv\cdot\grad)\Temp
		- \DtS\,\grad\Temp\,.
\end{align}
The derivatives of~$\DtS$ and~$\divPpv$ are then expanded in the usual manner
for~$\vv$-independent quantities as
\begin{equation}
	\Liouv\DtS = \Dt{\DtS} + \cv\cdot\grad\DtS,\qquad
	\Liouv\divPpv = \Dt{\divPpv} + \cv\cdot\grad\divPpv.
\end{equation}

The first term on the right-hand side of~\eqref{eq:fdistt2eq} becomes
\begin{align}
\Liouv \fdistt_1
	&= -\frac{1}{\Pra}\,
		\bigl(\RTv(\cnuv\cdot\divPpv + \cnuv\cdot\grad\ln\pres)
		- 2\cnuv\,\cnuv:\trost
		- \cnu^2(\DtS + \cv\cdot\grad\ln\Temp)\bigr)
		\cv\cdot\grad\ln{\Temp}\nonumber\\
	\pheq - \frac{1}{\Pra}\l(\cnu^2
			- \frac{5}{2}\r)\l(\half\RTv^2(\divPpv + \grad\ln\pres)
		- \cv\cdot\rost + \cv\times\vortv\r)
		\cdot\grad\ln{\Temp}\nonumber\\
	\pheq - \frac{1}{\Pra}\,\frac{1}{\Temp}\l(\cnu^2
			- \frac{5}{2}\r)\cv\cdot\l(\Dt{\grad\Temp}
			+ \cv\cdot\grad\grad\Temp
		+ \tfrac{2}{3}\,\div\uv\,\grad\Temp
		- \grad\ln\Temp\,(\cv\cdot\grad)\Temp
		- \DtS\,\grad\Temp\r)\nonumber\\
	\pheq - 4\cnuv\,\l(\half\RTv(\divPpv + \grad\ln\pres)
		- \cnuv\cdot\trost + \cnuv\times\vortv
		- \half\,\cnuv(\DtS + \cv\cdot\grad\ln\Temp)\r):\trost
		\nonumber\\
	\pheq - 2\cnuv\,\cnuv:\bigl(\Dt{\trost} + \cv\cdot\grad\trost\bigr)
		+ \frac{1}{\Pra}\,\l(\half\RTv^2(\divPpv + \grad\ln\pres)
		- \cv\cdot\rost + \cv\times\vortv\r)\cdot\divPpv\nonumber\\
	\pheq + \frac{1}{\Pra}\,\cv\cdot\l(\Dt\divPpv
		+ \cv\cdot\grad\divPpv\r)
		- \l(\cnu^2 - \frac{3}{2}\r)
		\l(\Dt{\DtS} + \cv\cdot\grad\DtS\r)\nonumber\\
	\pheq - \bigl(\RTv\cnuv\cdot(\divPpv + \grad\ln\pres)
		- 2\cnuv\,\cnuv:\trost
		- \cnu^2(\DtS + \cv\cdot\grad\ln\Temp)\bigr)\DtS
	\label{eq:Liouvfdistt1overrelt}\ .
\end{align}

Solving for~$\fdistt_2$ is now a matter of collecting the terms
of~\eqref{eq:quadterm},~\eqref{eq:Liouvrelt}
and~\eqref{eq:Liouvfdistt1overrelt}.  We then follow the same prescription as
in Section~\ref{sec:firstorder} and assign a relaxation time~$\relt$ to the
terms in~$\fdistt_2$ that are even in~$\cv$, and~$\relt/\sigma$ to the odd
terms, to account for the different relaxation rates of momentum and energy.
Once this is done, we may verify, in an analogous manner to the procedure of
Section~\ref{sec:matching1}, that the matching conditions are satisfied
by~$\fdist_2$ to the order of the approximation.  This is a lengthy, but
straightforward computation that we omit here.

\subsection{Pressure Tensor and Heat Flux}
\label{sec:prestheatf2}

As in Section~\ref{sec:prestheatf1}, the second-order pressure tensor and heat
flux are obtained by inserting the solution~$\fdistt_2$ of
Section~\ref{sec:distr-funct2} into~\eqref{eq:PrestQvdef}.  On performing the
required integrations we find, for the pressure tensor,
\begin{multline}
\Prest^{(2)} = \pres\,\relt^2
\biggl\lgroup
\Bigl\lgroup\Dt{(2\trost + \DtS\,\unitI)}
		- 2\Trlessw{\grad\uv\cdot(2\trost + \DtS\,\unitI)}\Bigr\rgroup
	+ \frac{\RTv^2}{\Pra\Temp}\Trlessw{\grad\grad\Temp}
	+ \frac{\sT\,\RTv^2}{\Pra\Temp^2}\,\Trlessw{\grad\Temp\,\grad\Temp}\\
	+ 8\,\Trlessw{\trost\cdot\trost}
	+ \frac{4}{3}(\tfrac{7}{2} - \sT)\,\div\uv\,\trost
	+ \frac{5\RTv^2}{6\Pra\,\Temp}\lapl\Temp\,\unitI
	+ \frac{5\sT\,\RTv^2}{6\Pra\Temp^2}|\grad\Temp|^2\,\unitI
	+ \frac{4}{3}\,\trost:\trost\,\unitI\\
	+ 2(\sT - 5)\,\DtS\,\trost
	+ \frac{\RTv^2}{2\Pra}\,(17-3\sT)\,\Trlessw{\divPpv\,\grad\ln\Temp}
	+ \frac{\RTv^2}{\Pra}\,\Trlessw{\divPpv\,\grad\ln\pres}
	+ \frac{\RTv^2}{\Pra}\,\Trlessw{\divPpv\,\divPpv}
	- \frac{\RTv^2}{\Pra}\,\Trlessw{\grad\divPpv}\\
	+ \sT\,\DtS^2\,\unitI
	+ \frac{\RTv^2}{3\Pra}\,\divPpv\cdot\grad\ln\pres\,\unitI
	+ \frac{\RTv^2}{\Pra}\,(\tfrac{5}{2} - \tfrac{1}{3}\,\sT)
		\,\divPpv\cdot\grad\ln\Temp\,\unitI\\
	\frac{1}{3}\,(5 - 2\sT)\,\div\uv\,\DtS\,\unitI
	+ \frac{\RTv^2}{3\Pra}\,\divPp^2\,\unitI
	- \frac{5\RTv^2}{6\Pra}\,\div\divPpv\,\unitI
\biggr\rgroup\, ,
	\label{eq:Prest2}
\end{multline}
and for the heat flux,
\begin{multline}
\Qv^{(2)} = \tfrac{1}{4}\,\pres\,\RTv^2\,\relt^2\,\Pra^{-1}
\Bigl\lgroup
	5\,\Pra^{-1}\,\frac{1}{\Temp}
		\bigl(\Dt{\grad\Temp} - \grad\uv\cdot\grad\Temp\bigr)
	+ 14\l(\sT + (1+\Pra^{-1})\r)\trost\cdot\grad\ln\Temp\\
	+ \tfrac{10}{3}\,\Pra^{-1}(\tfrac{7}{2} - \sT)\div\uv\,
		\frac{\grad\Temp}{\Temp}
	+ 14\,\div\trost
	- 4\,\trost\cdot\grad\ln\pres\\
	- 5\Pra^{-1}(\Dt{\divPpv} - \grad\uv\cdot\divPpv\bigr)
	+ 10\grad\DtS
	- 2(2 + 7\Pra^{-1})\,\divPpv\cdot\trost\\
	- 5\bigl(1 + (\sT + 1)\Pra^{-1}\bigr)\,\DtS\,\divPpv
	- \Pra^{-1}\tfrac{5}{3}(5 - 2\sT)\,\div\uv\,\DtS\,\divPpv\\
	+ 5\l((\tfrac{3}{4} + 2\sT) + \sT\,\Pra^{-1}\r)\,
		\DtS\,\grad\ln\Temp - 5\,\DtS\,\grad\ln\pres
\Bigr\rgroup.
	\label{eq:Qv2}
\end{multline}
A circle on top of a bar
means that the entire term
under the bar should be symmetrized and rendered traceless as implied by the
notation of~\eqref{eq:trlessdef}.

We may expand the second-order pressure-tensor~\eqref{eq:Prest2} and heat
flux~\eqref{eq:Qv2} in $\tau$, to recover the same terms as one obtains using
the Chapman-Enskog approach, that is, what is usually called the BGK--Burnett
equations.  In doing this, we discard terms of order $\tau^3$, include the
second-order contributions from~$\Prest^{(1)}$ and~$\Qv^{(1)}$ and treat the
Prandtl number in the same way in both cases.  When we put in a Prandtl number
of $2/3$, and discard cubic terms, we recover the conventional Burnett result
for the Boltzmann collision operator, which is a justification for the Prandtl
number fix discussed in Section~\ref{sec:prandtl}.  To illustrate the nature
of the new form for the pressure tensor and heat flux, we treat the case of
ultrasonic sound waves in the next section.

\section{Ultrasonic Waves}
\label{sec:ultrasonic}

It is known experimentally~\cite{Meyer1957} that, when the frequency of a
sound wave becomes comparable with the collision frequency of particles, both
the \NS\ and Burnett equations give poor accounts of the dispersion relations.
In Refs.~\cite{Chen2000,Chen2001b,Chen2002} the present theory was found to
give very good agreement with the data for phase velocities, in first order,
so that it is not possible to do much better with higher-order terms to within
experimental error.  However, the first-order theory predicts no damping of
the waves at large Knudsen number.  One of our motivations for this work was
to see if we could capture the experimental damping in higher order.  As we
shall see, the second-order theory does better than the first order theory for
a Knudsen number up to about 8.  The price for this seems to be negative
damping for Knudsen numbers above 8.  This instability of the homogenous state
appears also in the Burnett theory and it remains to be seen whether any of
the regularizations proposed for that case will work for our equations.

We start by expanding the macroscopic variables about an equilibrium rest
state ($\uv_0\equiv0$),
\begin{equation}
	\dens = \dens_0(1 + \ddens), \quad
	\Temp = \Temp_0(1 + \dTemp), \quad
       p = p_0(1 + \dpres) \, ,
\end{equation}
and  find from the equation of state~\eqref{state} that $\dpres = \ddens +
\dTemp.$ The linearized equations of
motion~\eqref{eq:denseq}--\eqref{eq:Tempeq} are then
\begin{gather}
	\pdt\ddens + \div\uv = 0,\label{eq:denseqlin}\\
	\dens_0\pdt\uv + \div(\Prest^{(0)} + \Prest^{(1)} + \Prest^{(2)}) = 0,
	\label{eq:ueqlin}\\
	\tfrac{3}{2}\pres_0\pdt\dTemp + \pres_0\div\uv
		+ \div(\Qv^{(1)} + \Qv^{(2)}) = 0,
		\label{eq:Tempeqlin}
\end{gather}
where the linearized pressure tensor in each order is
\begin{equation}
	\Prest^{(0)} = \pres_0\,\dpres\,\unitI;\qquad
	\Prest^{(1)} = -\visc_0\bigl(\grad\uv + \transp{\grad\uv}
		+ \pdt\dTemp\,\unitI\bigr);
\end{equation}
\begin{multline}
\Prest^{(2)} = \visc_0\relt_0\Bigl\lgroup
	(1 + \Pra^{-1})\,\pdt(\grad\uv + \transp{\grad\uv})
	+ \frac{1}{\Pra}\,\pdt\div\uv\,\unitI\\
	+ \pdt^2\dTemp\,\unitI
	+ \frac{\RTv_0^2}{\Pra}\,(\grad\grad\dTemp
		+ \tfrac{1}{2}\,\lapl\dTemp\,\unitI)
	+ \frac{\RTv_0^2}{\Pra}\,
		(\grad\grad\dpres + \tfrac{1}{2}\,\lapl\dpres\,\unitI)
\Bigr\rgroup,
\end{multline}
and the linearized heat flux is
\begin{equation}
	\Qv^{(1)} = -\tcond_0\Temp_0\grad(\dTemp+\dpres)
	- \tfrac{5}{2}\,\visc_0\,\Pra^{-1}\,\pdt\uv\,;
\end{equation}
\begin{equation}
\Qv^{(2)} =
\tfrac{1}{4}\,\visc_0\,\RTv_0^2\,\relt_0\,\Pra^{-1}
\Bigl\lgroup
	\frac{10}{\Pra\RTv_0^2}\,\pdt^2\uv
	+ 5\bigl(2 + \Pra^{-1}\bigr)\,\pdt{\grad\dTemp}
	+ \frac{5}{\Pra}\,\pdt\grad\dpres
	+ 7\,\lapl\uv + 9\grad(\div\uv)
\Bigr\rgroup.
\end{equation}
The viscosity~$\visc$ and the thermal conductivity~$\tcond$ are defined by
Eq.~\eqref{eq:diff}.  The derivatives of~$\relt$ do not enter the
linearization, as we see from the absence of the parameter~$\sT$
defined in~\eqref{xizeta}. 

We take the divergence of the~$\uv$ equation~\eqref{eq:ueqlin}, and use the
continuity equation~\eqref{eq:denseqlin} to eliminate~$\div\uv$.  After some
manipulation we find
\begin{multline}
\relt_0^2\,\pdt^2\ddens =
\cs^2\relt_0^2\,\lapl(\ddens + \dTemp)
+ \frac{\relt_0^3}{\cs^2}\,\bigl(2\pdt\lapl\ddens - \pdt\lapl\dTemp\bigr)\\
+ \frac{\relt_0^4}{\cs^2}\Bigl\lgroup
	-(2 + 3\Pra^{-1})\,\pdt^2\lapl\ddens
	+ \pdt^2\lapl\dTemp
	+ 6\cs^2\Pra^{-1}\,\blapl\dTemp
	+ 3\cs^2\Pra^{-1}\,\blapl\ddens
\Bigr\rgroup,
\label{eq:ddenseq}
\end{multline}
where
\begin{equation}
	\cs^2 \ldef \frac{\pres_0}{\dens_0}, \qquad
	\kvisc_0 \ldef \frac{\visc_0}{\dens_0}
\end{equation}
are respectively the isothermal sound speed and kinematic viscosity, in terms
of which~\hbox{$\relt_0 = \visc_0/\pres_0 = \kvisc_0/a^2$}.
The equation for the temperature
perturbation~\eqref{eq:Tempeqlin} becomes
\begin{multline}
\tfrac{3}{2}\,\relt_0\,\pdt\dTemp - \relt_0\,\pdt\ddens
- \tfrac{5}{2}\cs^2\relt_0^2\,\Pra^{-1}\lapl(2\dTemp + \ddens)
	+ \tfrac{5}{2}\,\relt_0^2\,\Pra^{-1}\,\pdt^2\ddens\\
+ \tfrac{1}{2}\,\relt_0^3\,\cs^2\,\Pra^{-1}
\Bigl\lgroup
	-\frac{5}{\Pra\cs^2}\,\pdt^3\ddens
	+ 10\bigl(1 + \Pra^{-1}\bigr)\,\pdt{\lapl\dTemp}
	+ (5\Pra^{-1} - 16)\,\pdt\lapl\ddens
\Bigr\rgroup = 0.
\label{eq:dTempeq}
\end{multline}

If we use~$\relt_0$ as unit of time and~$\cs\relt_0$ as unit of length, 
but do not change the meaning of the symbols used otherwise, we can
rewrite~\eqref{eq:ddenseq} and~\eqref{eq:dTempeq} as 
\begin{equation}
\l(\pdt^2 - \lapl - 2\pdt\lapl + (2 + 3\Pra^{-1})\,\pdt^2\lapl
	- 3\Pra^{-1}\,\blapl\r)\ddens
+ (\pdt - 1 - \pdt^2 - 6\Pra^{-1}\lapl)\lapl\dTemp = 0,
\label{eq:ddenseqd}
\end{equation}
and
\begin{multline}
\l(3\pdt - 10\Pra^{-1}\lapl + 10\Pra^{-2}(\Pra + 1)\,\pdt\lapl\r)\dTemp\\
	+ \Pra^{-1}\l(5\pdt^2 - 5\lapl - 2\Pra\pdt
	- 5\Pra^{-1}\pdt^3
	+ \Pra^{-1}(5 - 16\Pra)\,\pdt\lapl\r)\ddens
= 0.
\label{eq:dTempeqd}
\end{multline}
Taking the density and
temperature perturbations~$\ddens$ and~$\dTemp$ proportional
to~$\exp(\mathrm{i}(kx-\omega\time))$ we obtain a relation between~$k$
and~$\omega$, assuming neither of~$\ddens$ and~$\dTemp$ vanishes identically.
The number~\hbox{$\varepsilon = \omega\relt_0$} is dimensionless and can be
regarded as the Knudsen number, the ratio of the relaxation time ($\relt_0$)
to the macroscopic time ($1/\omega$).  The approximation involved in a
\CE-type expansion get worse as~$\omega\tau_0$ becomes large.

\begin{figure}
\psfrag{vir}[Bc][Bc]{\raisebox{2ex}{\Large $\mathrm{Re}\,(k/\omega)$}}
\psfrag{bi}{\raisebox{-1.5ex}{\Large \hspace{-1em} $1/\omega\relt_0$}}
\psfrag{Navier-Stokes}{\large Navier-Stokes}
\psfrag{Burnett      }{\large Burnett}
\psfrag{First-order  }{\large First-order}
\psfrag{Second-order }{\large Second-order}
\psfrag{Experiment   }{\large Experiment}
\centerline{\includegraphics[width=\textwidth]{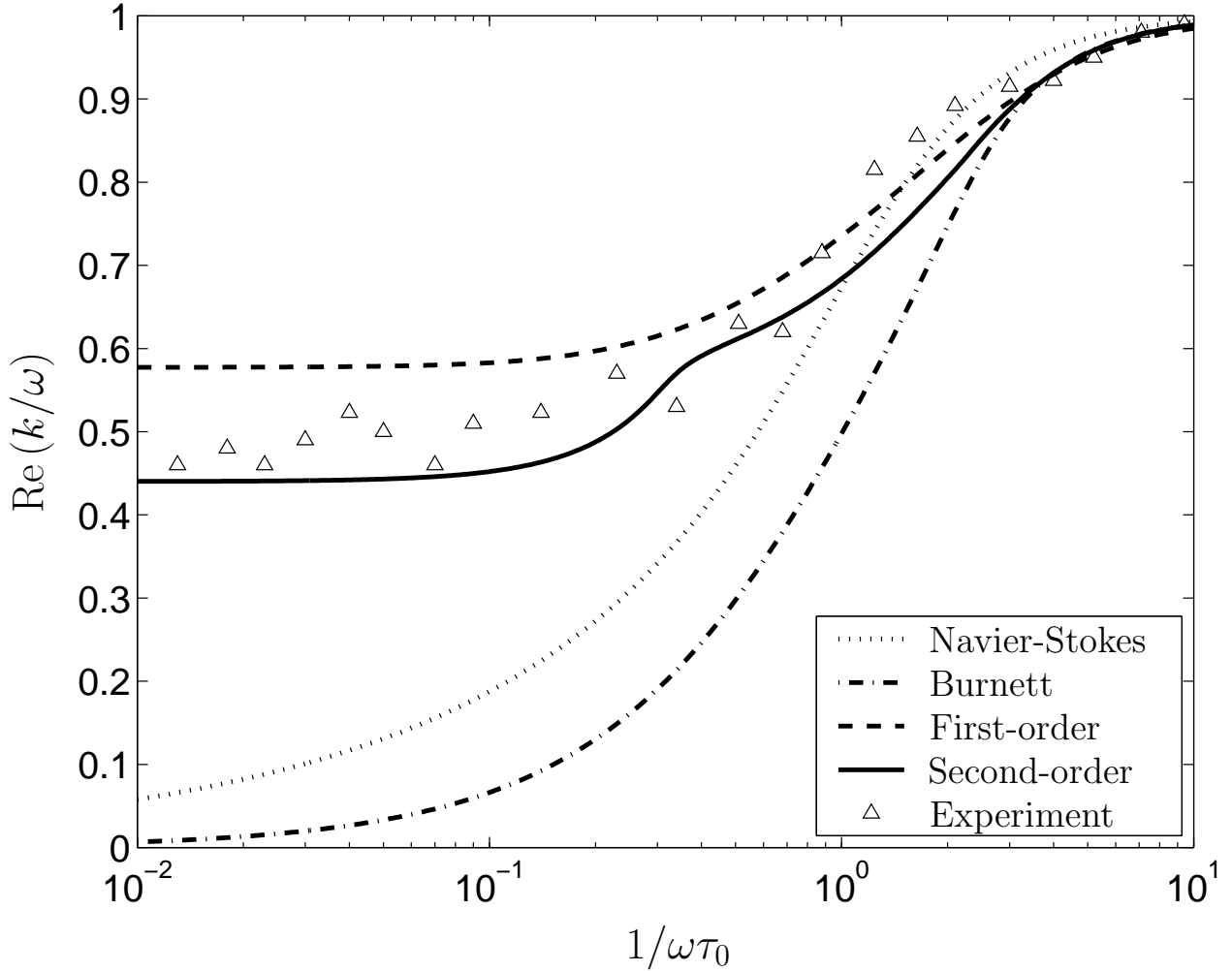}}
\caption{The real part of the inverse phase speed, $k/\omega$,
  versus~$1/\omega\relt_0$. The phase speed is given in units of the adiabatic
  sound speed,~$(\gamma\pres_0/\dens_0)^{1/2}$.  The first- and second-order
  dispersion relations of this paper (with~$\Pra=2/3$) are compared to \NS,
  Burnett, and the experiment of Meyer and Sessler~\cite{Meyer1957}.}
\label{fig:alphar}
\end{figure}

\begin{figure}
\psfrag{vii}[Bc][Bc]{\raisebox{1.5ex}{\Large $\mathrm{Im}\,(k/\omega)$}}
\psfrag{bi}{\raisebox{-1.5ex}{\Large \hspace{-1em} $1/\omega\relt_0$}}
\psfrag{Navier-Stokes}{\large Navier-Stokes}
\psfrag{Burnett      }{\large Burnett}
\psfrag{First-order  }{\large First-order}
\psfrag{Second-order }{\large Second-order}
\psfrag{Experiment   }{\large Experiment}
\centerline{\includegraphics[width=\textwidth]{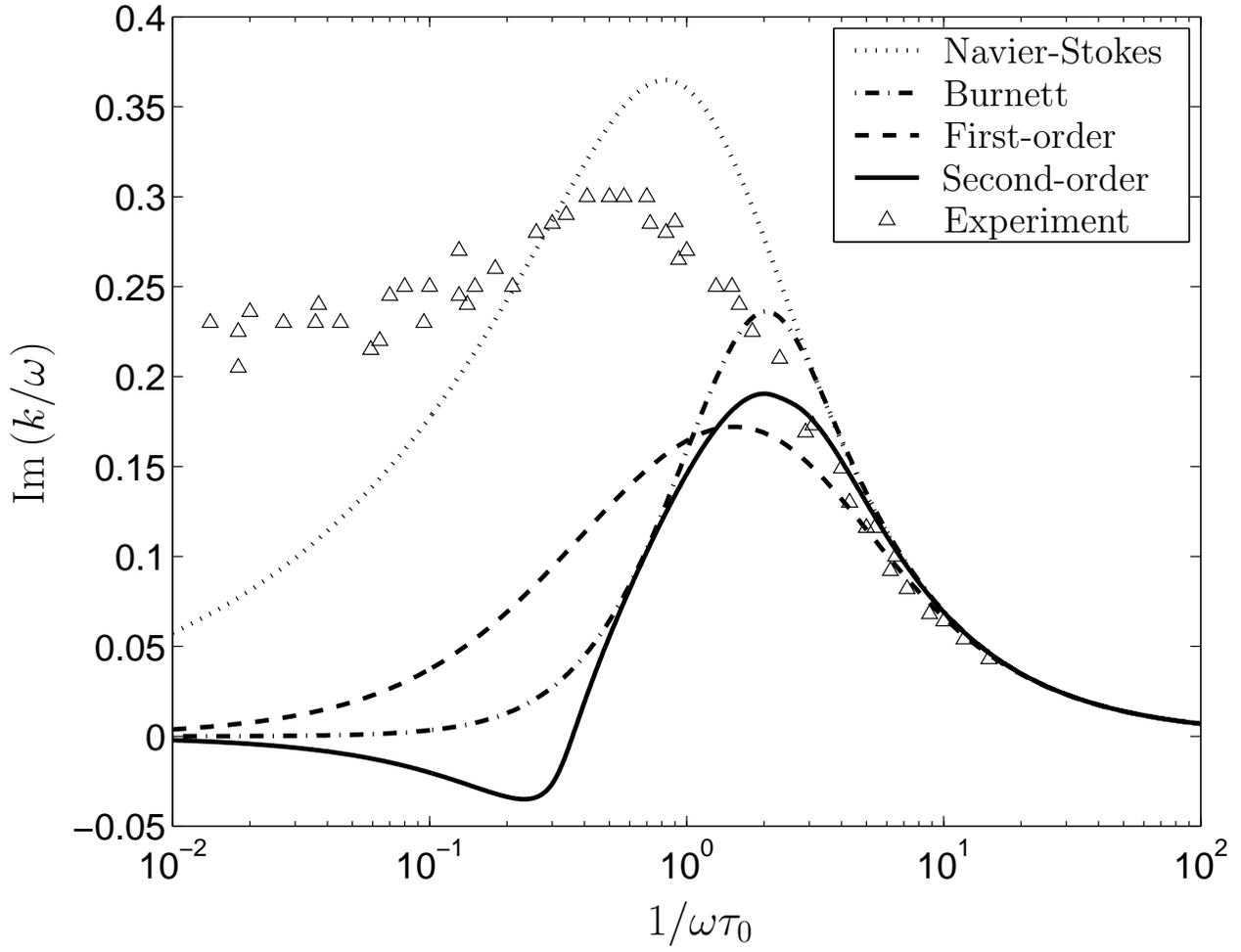}}
\caption{As for Fig.~\ref{fig:alphar}, but for the imaginary part of the
  inverse phase speed.}
\label{fig:alphai}
\end{figure}

To compare to experimental results, we plot the inverse phase speed~$k/\omega$
as a function of~$1/\omega\relt_0$.  Figure~\ref{fig:alphar} shows the real
part, and Fig.~\ref{fig:alphai} the imaginary part (damping).  The phase speed
is in units of the adiabatic sound speed,~$(\gamma\pres_0/\dens_0)^{1/2}$,
where~$\gamma=5/3$.  We used a Prandtl number of $2/3$ and so compare with the
N--S and Burnett results from the Boltzmann equation for Maxwell molecules.

\section{Discussion}
\label{sec:discussion}

According to Grad~\cite{Grad1963} and Uhlenbeck~\cite{Uhlenbeck}, the Hilbert
expansion of $f$ in terms of mean flight times is asymptotic and not
convergent.  For this reason, we have avoided using an iterative procedure
like those of Chapman and of Enskog.  In their ways of deriving the fluid
equations~\cite{ChapmanCowling2nd}, it is the practice to use lower-order
approximations to simplify the current approximation.  In particular, the \CE\
approach uses the Euler equations to simplify the calculation of the fluid
equations at first order, and the \NS\ equations at second order.  Here, we do
not recycle lower-order results in this way and so we obtain equations at a
given order of accuracy $N$ that differ from the \CE\ results by terms
$\Order{\relt^{N+1}}$.  In other words, our results agree asymptotically with
those of the \CE\ procedure but, when we apply the two theories at the edges
of their validities, we see differences in their quantitative predictions.

The effect of this difference in approach is seen starkly when we take the
trace of the stress tensor, $\mathbb{P}$.  From the definition of the pressure
tensor, its trace is seen to be
\begin{equation}
\label{extrace}
\Tr{\mathbb{P}}=3p
\end{equation}
while our $N$th approximation to the stress tensor, $\mathbb{P}_N$\,, has a
trace of $3p+\mathcal{O}(\tau^{N+1})$.  It appears to be a common
misconception that it is required to make the $\mathcal{O}(\tau^{N+1})$ error
in the trace of the pressure tensor vanish identically.  However, we are
perfectly at liberty to retain terms $\mathcal{O}(\tau^{N+1})$ in a theory
that is intended to be good to $\mathcal{O}(\tau^{N})$.  Indeed, we suggest
that, in the present instance, it might even be desirable to retain the extra
terms in $\mathbb{P}$.

In the approach described here, we have in the order $N$, an approximation for
the pressure tensor with an error of order $\tau^{N+1}$, so that we may write
$\mathbb{P} = \mathbb{P}_N + \mathcal{O}(\tau^{N+1})$.  Let us put this as
\begin{equation}
\mathbb{P} = p \mathbb{I} + \mathbb{A}  \tau^{N}+ \mathcal{O}(\tau^{N+1})
\end{equation}
where $\mathbb{A}$ is the tensor we need to find to complete our approximation.
When we take the trace of this equation, we find the requirement that
\begin{equation}
\Tr{\mathbb{A}} + \mathcal{O}(\tau) = 0.
\label{A}
\end{equation}

Iterative procedures like \CE\ theory force the trace of $\mathbb{A}$ to
vanish exactly and, on that account, (\ref{A}) cannot be satisfied unless
$\tau$ is truly infinitesimal.  But, in our approach, (\ref{A}) is not an
embarrassment since it points a way to increase the domain of validity of our
approximation.  By making a suitable choice of $\mathbb{A}$ we might cancel
some of the remaining error in the trace of $\mathbb{P}$ if we but knew how to
arrange this.  As we have observed, the expansion procedure offers a natural
choice for $\mathbb{A}$ and the problem is then to test its accuracy.

As in most physics problems, we prefer to let the choice among the theories be
made by the experiments, which do favor our approach over the standard \CE\
treatment.  Indeed, with the Prandtl number adjustment, our results from the
relaxation model seem to be preferred to those from the Boltzmann equation in
the \CE\ approach.  We shall return to the discussion of our approach to the
Boltzmann equation in another place.  Here, we have remained in the relatively
simple case of the relaxation model whose results seem as good as those with
the Boltzmann model once the Prandtl number is adjusted (see also the
discussion of the so-called BGK--Burnett equations in~\cite{Agarwal2001}). 

In the study of ultrasound, the propagation speed we predict agrees well with
the experimental findings at all Knudsen numbers.  However, our results for
the damping of ultrasound are good (as judged by experiment) only for Knudsen
number up to order unity.  Since our second-order results are an improvement
over the first order ones, we may hope for improvement as higher orders are
achieved.  But in practice, going to higher order than second would seem to be
prohibitively complicated, even if the calculations could be performed.  The
problem is then what ought to be the approach to the damping of sound waves in
the limit of very long mean flight times of particles.

One clue is that for moderate Knudsen numbers we have some negative damping
for a small range of flight times.  This is reminiscent of the instability
found in the Burnett equations and Jin and Slemrod~\cite{Jin2001,Jin2001b}
treat this problem with a regularization procedure that involves manipulation
of high order terms, as Rosenau has also done~\cite{Rosenau1989}.  Such
procedures have also proved effective in problems of photon
transport~\cite{Stress2000}.  Similar procedures may be used to ameliorate our
equations and we shall return to that possibility elsewhere.  But there is an
interesting aspect to this.  The various regularization procedures being
introduced to improve the continuum equations generally involve the
introduction of higher order terms.  Here, we are making a plea for first
looking more carefully at the higher order terms that are discarded by
standard procedures.  However, we wish to conclude with mention of a final
issue.

Stubbe~\cite{Stubbe1990} has argued that the dynamics of rarefied gases cannot
be well treated macroscopically and a that microscopic approach is needed as
in the case of plasma physics where no macroscopic approach has produced
Landau damping~\cite{Landau1946,Stubbe1999}.  On the other hand, it may be
that a suitable resummation of the results of lower orders would resolve the
issue.  In any case, it appears that the study of the damping of ultrasound
waves is going to involve serious complications.

\begin{acknowledgments}

We are happy to acknowledge that some of the work reported here was carried
out during our participation in the Summer Program in Geophysical Fluid
Dynamics at the Woods Hole Oceanographic Institution.  J-LT thanks David
Maurer for pointing out a useful reference.  EAS thanks S. Jin, M. Slemrod and
J. Dufty for interesting discussions.

\end{acknowledgments}

\appendix

\clearpage

\section{Recovering Burnett}

As we have said, the Chapman-Enskog procedure eliminates terms of higher order
than does the current procedure.  This elimination can be performed at various
stages and, when we eliminate those same terms from our equations, we recover
the Chapman-Enskog type of results in the N--S and Burnett levels of
approximation.  More surprsingly perhaps is that when we set the Prandtl
number to $2/3$ the same elimination of higher order terms gives us the
Burnett equations with exactly the same coefficients as is found with the
Boltzmann equation for hard spheres.  Here, we sketch this reduction.

Equations~\eqref{eq:Prest2}--\eqref{eq:Qv2} contain terms in~$\DtS$
and~$\divPpv$, which are formally of order~$\relt$.  We may thus write
\begin{multline}
\Prest^{(2)} = \pres\,\relt^2
\biggl\lgroup
2\Bigl\lgroup\Dt{\trost} - 2\Trlessw{\grad\uv\cdot\trost}\Bigr\rgroup
	+ \frac{\RTv^2}{\Pra\Temp}\Trlessw{\grad\grad\Temp}
	+ \frac{\sT\,\RTv^2}{\Pra\Temp^2}\,\Trlessw{\grad\Temp\,\grad\Temp}
	+ 8\,\Trlessw{\trost\cdot\trost}\\
	+ \frac{4}{3}(\tfrac{7}{2} - \sT)\,\div\uv\,\trost
	+ \frac{5\RTv^2}{6\Pra\,\Temp}\lapl\Temp\,\unitI
	+ \frac{5\sT\,\RTv^2}{6\Pra\Temp^2}|\grad\Temp|^2\,\unitI
	+ \frac{4}{3}\,\trost:\trost\,\unitI
\biggr\rgroup + \Order{\relt^3}
	\label{eq:Prest2B}
\end{multline}
and
\begin{multline}
\Qv^{(2)} = \tfrac{1}{4}\,\pres\,\RTv^2\,\relt^2\,\Pra^{-1}
\Bigl\lgroup
	5\,\Pra^{-1}\,\frac{1}{\Temp}
		\bigl(\Dt{\grad\Temp} - \grad\uv\cdot\grad\Temp\bigr)
	+ 14\l(\sT + (1+\Pra^{-1})\r)\trost\cdot\grad\ln\Temp\\
	+ \tfrac{10}{3}\,\Pra^{-1}(\tfrac{7}{2} - \sT)\div\uv\,
		\frac{\grad\Temp}{\Temp}
	+ 14\,\div\trost
\Bigr\rgroup + \Order{\relt^3}.
	\label{eq:Qv2B}
\end{multline}
These are not the second-order pressure and heat flux of the Burnett
equation.  To recover the Burnett result, we need to expand to first order
in~$\relt$ the~$\DtS$ and~$\divPp$ terms appearing in the first order pressure
tensor and heat flux~\eqref{eq:PrestQv1}.

First we do this for~$\divPpv$, which enters the heat flux
in~\eqref{eq:divPp2}.  On inserting~\eqref{eq:PrestQv1} in~\eqref{eq:divPp2}
we find
\begin{equation}
 \divPpv = \frac{1}{\pres}\,\div(-2\pres\,\relt\,\trost) + \Order{\relt^2}
\end{equation}
which can be expanded to give
\begin{equation}
 \divPpv = -2\relt\,\div\trost
 - 2\relt\,\trost\cdot\grad\log\pres
 - 2\relt\pres\,\trost\cdot\grad\log\relt + \Order{\relt^2}.
\end{equation}
These terms, when multiplied by~$(5\relt/4\Pra)\,\pres\,\RTv^2$ and added
to~\eqref{eq:Qv2B}, agree with the Burnett expression for the second-order
heat flux derived using a Boltzmann collision operator (after
setting~$\Pra=2/3$).

Now for~$\DtS$, which enters the pressure tensor in~\eqref{eq:divPp2}.  After
inserting~\eqref{eq:PrestQv1} in~\eqref{eq:DtS2} we obtain
\begin{equation}
  \DtS
  = -\frac{2}{3\pres} (-2\pres\relt\trost) \div\uv
  + \div\l(-\tfrac{5}{4}\,\frac{\relt}{\Pra}\,\pres\,\RTv^2\grad\ln{\Temp}\r)
  + \Order{\relt^2};
\end{equation}
Upon multiplying this by~$-\pres\,\relt\,\unitI$, adding
to~\eqref{eq:Prest2B}, and setting~$\Pra=2/3$, we recover the Burnett
expression for the second-order pressure tensor appropriate for the Boltzmann
collision operator.  Thus our full expressions for the pressure tensor and
heat flux agree with Burnett to second-order, as they must since both theories
(ours and Burnett's) are second-order expansions.  The perfect agreement also
validates the Prandtl number fix discussed in Section~\ref{sec:prandtl}.


\end{document}